\documentclass[useAMS]{mn2e}
\usepackage{times}
\usepackage{graphics}

\begin{document}

\title[Lyman Break Galaxies and the Star Formation Rate at $z\approx
6$]{Lyman Break Galaxies and the Star Formation Rate of the Universe at
$z\sim 6$}

\author[E.~Stanway, A.~Bunker \& R.~McMahon]{Elizabeth
R.~Stanway,$^{1}$\thanks{email:ers@ast.cam.ac.uk}
Andrew~J.~Bunker\,$^{1}$ \& Richard G.~McMahon\,$^{1}$\\
$^{1}$\,Institute of Astrophysics, University of Cambridge,
Madingley Road, Cambridge, CB3\,0HA, UK}
\date{Accepted 2003 February 14. Received 2003 February 13; in original
form 2002 December 03.}

\maketitle

\begin{abstract} 
We determine the space density of UV-luminous star-burst galaxies at
$z\sim 6$ using deep {\em HST} ACS SDSS-$i'$ (F775W) and SDSS-$z'$
(F850LP) and VLT ISAAC $J$ and $K_s$ band imaging of the Chandra Deep
Field South. We find 8 galaxies and one star with $(i'-z')>1.5$ to a
depth of $z'_{\rm AB}= 25.6$ (an $8\,\sigma$ detection in each of the 3
available ACS epochs). This corresponds to an unobscured star formation
rate of $\approx 15\,h_{70}^{-2}\,M_{\odot}\,{\rm yr}^{-1}$ at $z=5.9$,
equivalent to $L^*$ for the Lyman break population at $z = 3-4$
($\Omega_{\Lambda}=0.7$, $\Omega_{M}=0.3$). We are sensitive to star
forming galaxies at $5.6\la z \la 7.0$ with an effective comoving volume
of $\approx 1.8\times10^5\,h_{70}^{-3}\,{\rm Mpc}^{3}$ after accounting
for incompleteness at the higher redshifts due to luminosity bias. This
volume should encompass the primeval sub-galactic scale fragments of the
progenitors of about a thousand $L^*$ galaxies at the current epoch. We
determine a volume-averaged global star formation rate of $(6.7 \pm 2.7)
\times 10^{-4}\,h_{70}\,M_{\odot}\,{\rm yr}^{-1}\,{\rm Mpc}^{-3}$ at
$z\sim 6$ from rest-frame UV selected star-bursts at the bright end of
the luminosity function: this is a lower limit because of dust
obscuration and galaxies below our sensitivity limit. This measurement
shows that at $z\sim 6$ the star formation density at the bright end is
a factor of $\sim 6$ times less than that determined by Steidel et al.\
(1999) for a comparable sample of UV selected galaxies at $z=3-4$, and
so extends our knowledge of the star formation history of the Universe
to earlier times than previous work and into the epoch where
reionization may have occurred.

\end{abstract} 

\section{Introduction}
\label{sec:intro}

How, when, and over what time-scale galaxies formed are questions at the
forefront of both observational and theoretical cosmology. In recent
years there has been tremendous progress in the study of galaxy
formation from an observational perspective. Spectroscopic samples of
star-forming galaxies have been detected at progressively higher
redshifts: $z\sim 1$ (Cowie et al.\ 1996); $z\sim 3$ (Steidel et al.\
1996); $z\sim 4$ (Hu \& McMahon 1996; Steidel et al.\ 1999); beyond
$z=5$ (e.g., Spinrad et al.\ 1998; Weymann et al.\ 1998; Hu, McMahon
\& Cowie 1999; Dawson et al.\ 2001; Lehnert \& Bremer 2003) and now up
to $z=6.6$ (Hu et al.\ 2002; Kodaira et al.\ 2003).

Complementing this ground-based spectroscopic work there have been a
number of programmes based on the Lyman break ``photometric redshift''
technique (e.g., Steidel, Pettini \& Hamilton 1995). {\em Hubble Space
Telescope (HST)} imaging with WPC2 and NICMOS has extended such studies
to fainter magnitudes than accessible through ground-based spectroscopy
(e.g., Madau et al.\ 1996; Dickinson et al.\ 2000; Thompson, Weymann \&
Storrie-Lombardi 2001). From the results of this work, great progress
has been made in reconstructing ``the star formation history of the
Universe'' -- the evolution of the comoving volume-averaged star
formation rate (as shown on the `Madau-Lilly' diagram: Lilly et al.\
1996; Madau et al.\ 1996) over the redshift range $0-4$.

Here we describe a survey that extends such work to $z\sim 6$.
A survey may be considered useful for selecting candidate star 
forming galaxies if it satisfies the following primary criteria:
(i) it can distinguish $z\sim 6$ galaxies from lower redshift objects;
(ii) it is deep enough to reach star formation rates characteristic of
this population; (iii) the volume is sufficient to encompass a statistically
meaningful sample.

In this paper we show that public data from the Advanced Camera for
Surveys (ACS, Ford et al.\ 2002) on {\em HST}, released as part of the
public Great Observatories Origins Deep Survey (GOODS -- Dickinson \&
Giavalisco 2002\footnote{see {\tt
http://www.stsci.edu/ftp/science/goods/}}) programme, is of sufficient
depth and volume to detect very high redshift galaxies, and that colour
selection should reject less distant objects. We use this data set to
examine the space density of UV-luminous starburst galaxies at $z\approx
6$. Observational evidence of the Gunn-Peterson effect in $z\sim 6$ QSOs
(Becker et al.\ 2001) suggests that this is the epoch of reionization,
where starlight may for the first time dominate the Universe -- although
recent results on the polarization of the cosmic microwave background
from the WMAP satellite indicate that the onset of ionization may have
been at even higher redshift (Kogut et al.\ 2003). Using deep imaging by
ACS through its SDSS-$i'$ and -$z'$ filters, we identify a sample of
objects which are likely to be Lyman break star forming galaxies at
$5.6<z<7.0$. From the UV continuum longward of $\lambda_{\rm
rest}=1216$\,\AA\ detected in the $z'$-band, we place a lower limit on
the global star formation rate at $z\sim 6$ and hence extend the
Madau-Lilly diagram to early epochs.

Throughout we adopt the now widely accepted `concordance' cosmology: a
$\Lambda$-dominated, flat Universe with $\Omega_{\Lambda}=0.7$,
$\Omega_{M}=0.3$ and $H_{0}=70\,h_{70} {\rm km\,s}^{-1}\,{\rm
Mpc}^{-1}$. The {\em HST}/ACS magnitudes in this paper are quoted in the
AB system (Oke \& Gunn 1983).

\section{Observations}
\label{sec:obs}

\subsection{The ACS data}

We have analysed the first three epochs of GOODS data from the Chandra
Deep Field South (CDFS) taken with {\em HST}/ACS\footnote{available from
{\tt ftp://archive.stsci.edu/pub/hlsp/goods/}}. Each epoch of data
comprises a half-orbit $i'$-band image taken through the F775W filter
(1040\,sec, split into 2 for cosmic ray rejection) and a full-orbit
$z'$-band image observed with the F850LP filter (2010\,sec, split into 4).
We use the v0.5 release of the reduced data.

The released data set has been pipeline-processed for dark and bias
subtraction and flat-fielding, and known hot pixels, bad columns and
other cosmetic defects flagged in a data quality file. The
post-pipeline images are sky-subtracted,``drizzled''(Fruchter \& Hook
2002) and corrected for geometric distortion. Cosmic rays were also
identified and masked.

We use the zero-points for AB magnitudes determined by the GOODS team:
${\rm mag}_{\rm AB} = {\rm zeropoint}-2.5\log_{10}({\rm
Count~rate} / {\rm s}^{-1}) $ where the zeropoints for $i'$-band
(F775W) and $z'$-band (F850LP) are $25.66$ and $24.92$ respectively. We
have corrected for the small amount of foreground Galactic extinction
toward the CDFS using the {\it COBE}/DIRBE \& {\it IRAS}/ISSA dust maps
of Schlegel, Finkbeiner \& Davis (1998). The optical reddening is
$E(B-V)=0.008$, equivalent to extinctions of $A_{F775}=0.017$ \&
$A_{F850LP}=0.012$.

\subsection{Construction of the catalogues}

Candidate selection for all objects in the field was performed using
version 2.2.2 of the SExtractor photometry package (Bertin \& Arnouts
1996). As we are searching specifically for objects which are only
securely detected in $z'$, with little or no flux in the $i'$-band,
fixed circular apertures $1\farcs0$ in diameter were trained in the
$z'$-image and the identified apertures were used to measure the flux at
the same spatial location in the $i'$-band image, running SExtractor in
two-image mode. For object identification, we demanded at least 5
contiguous pixels above a threshold of $2\sigma$ per pixel ($0.01\,{\rm
counts\,pixel}^{-1}\,{\rm s}^{-1}$) on the drizzled data (with a pixel
scale of 0\farcs05~pixel$^{-1}$). From the output of SExtractor, we
create a sub-catalogue of all objects brighter than $z'_{\rm AB}<25.6$\,mag,
which corresponds to an $8\,\sigma$ detection in each epoch.

In the compilation of the catalogues a number of `figure-eight' shaped
optical reflections and diffraction spikes near bright stars were
masked, as was the gap between the two ACS CCDs and its vicinity in each
tile. This means that the survey area is non-contiguous.

Excluding the masked regions, the total survey area in the 15 tiles of
CDFS is 146\,arcmin$^{2}$. In this region, we detect 10728 sources in
the $z'$-band brighter than $z'_{\rm AB}=25.6$ (our $8\,\sigma$ cut for the
F850LP image from each individual epoch). The IRAF `Artdata' package
was used to create artificial objects of known magnitude and we
confirmed that we recover $>$95 per cent of such objects at the
magnitude limit of our catalogue.

\section{Candidate $z>5.6$ galaxies}
\label{sec:candidates}

\subsection{Redshift Discrimination} 
\label{sec:RedshfitDiscrim}

\begin{figure}
\resizebox{0.48\textwidth}{!}{\includegraphics{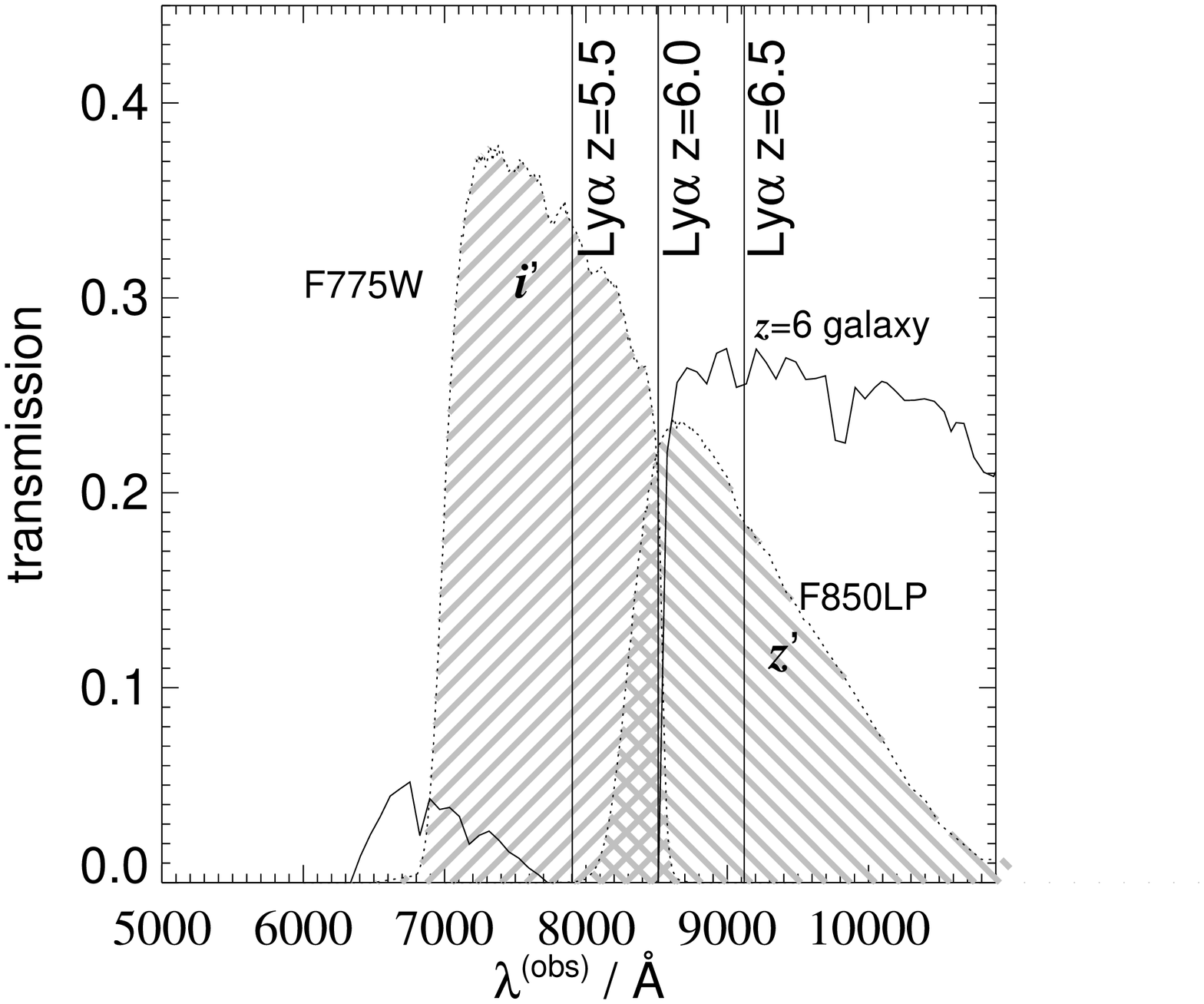}}
\caption{The ACS-$i'$ and -$z'$ bandpasses overplotted on
the spectrum of a generic $z=6$ galaxy (solid line), illustrating the
Lyman break technique.}
\label{fig:filters}
\end{figure}

In order to select $z>6$ galaxies, we use the Lyman break technique
pioneered at $z\sim 3$ from the ground by Steidel and co-workers and
using {\em HST} by Madau et al.\ (1996). The usual Lyman break technique
at $z\sim 3-4$ involves three filters: one below the Lyman limit ($\rm
\lambda_{rest}=912$\,\AA ), one in the Lyman forest region and a third
longward of the Lyman-$\alpha$ line ($\rm \lambda_{rest}=1216$\,\AA). At
higher redshifts, we use just two filters, since at $z\sim 6$ the
integrated optical depth of the Lyman-$\alpha$ forest is $\gg 1$. Hence
the continuum break at the wavelength of Lyman-$\alpha$ is large, as is
illustrated in Figure~\ref{fig:filters}, and the shortest-wavelength
filter below the Lyman limit becomes redundant. Thus provided one works
at a sufficiently-high signal-to-noise ratio, $i'$-band drop-outs can be
safely identified through detection in a single redder band (i.e.,
SDSS-$z'$). This has been elegantly shown by Fan and the SDSS
collaboration in the detection of $z\sim 6$ quasars using the $i'$-
and $z'$-bands alone (Fan et al.\ 2001). The sharp sides of the SDSS
filters help in clean selection using the photometric redshift
technique.

\begin{figure}
\resizebox{0.48\textwidth}{!}{\includegraphics{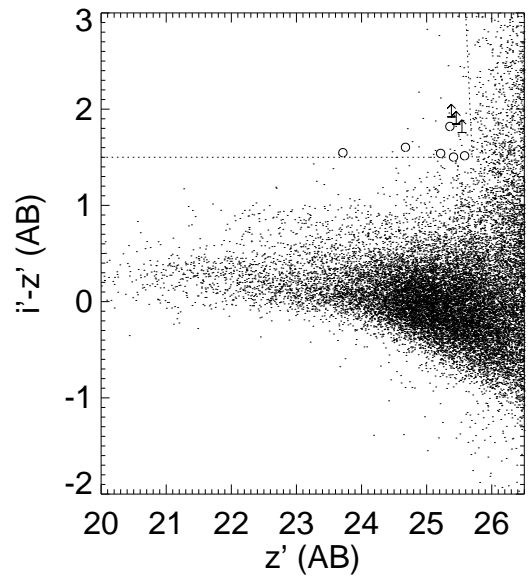}}
\caption{A colour-magnitude diagram for our data with the $z'$-band limit 
of $z'_{\rm AB}<25.6$ and the colour-cut
of $(i'-z')_{\rm AB}=1.5$ shown (dashed lines). Candidate objects are marked
with circles, or $3\,\sigma$ lower-limits on the $(i'-z')$ colour where
objects are undetected in $i'$.
Note: A catalogue based on a simple colour cut is contaminated by stars, 
EROs and wrongly identified extended objects as described in the text.}
\label{fig:izzcolmag}
\end{figure}

In Figures~\ref{fig:tracks}\,\&\,\ref{fig:izzcolmag} we illustrate how
the use of a colour cut of $(i'-z')_{\rm AB}>1.5$ can be used to select
objects with $z>5.6$, and we have 9 at $z'_{\rm AB}<25.6$. Objects
satisfying our selection criteria in any of the three epochs were
gathered into a preliminary catalogue. The flux of these objects in the
three epochs was averaged for each waveband, and those objects still
satisfying the criteria $z'_{\rm AB}<25.6$ and $(i'-z')_{\rm AB}>1.5$ in the
combined data set were selected. A number of objects appearing in
catalogs generated by the SExtractor software were identified by eye as
different regions of the same extended source (visible in the ACS F606W
$v$-band or F775W $i'$-band images) and eliminated from the final
selection. The 9 remaining objects satisfying our selection criteria are
detailed in Table~1.

\subsection{Interlopers at Different Redshifts}
\label{sec:interlopers}

\begin{figure}
\resizebox{0.48\textwidth}{!}{\includegraphics{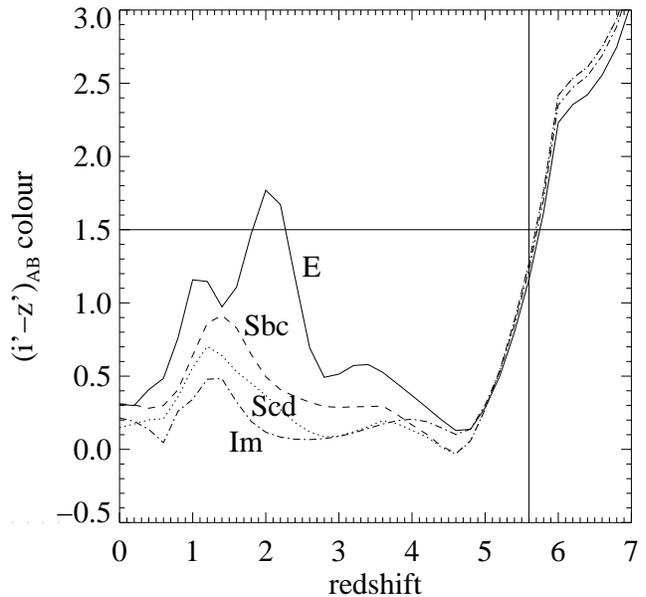}}
\caption{Model colour-redshift tracks for galaxies with non-evolving
stellar populations (from Coleman, Wu \& Weedman 1980 template spectra).
The `hump' in $(i'-z')$ colour seen at $z\approx 1-2$ is due
to the 4000\,\AA\ break redshifting beyond the $i'$-filter. }
\label{fig:tracks}
\end{figure}

\begin{figure}
\resizebox{0.48\textwidth}{!}{\includegraphics{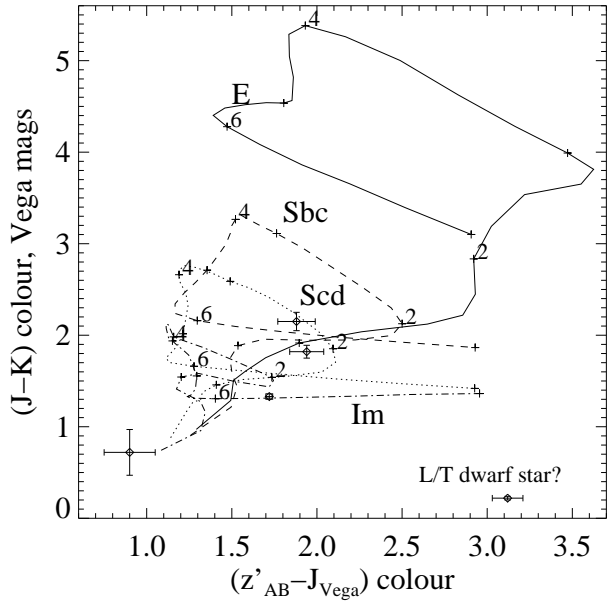}}
\caption{Model $(z'-J)$ -- $(J-K)$ tracks for galaxies with non-evolving
stellar populations. Redshift intervals are marked, and we plot
the colours of our $(i'-z')_{\rm AB}>1.5$ objects for which near-infrared
data is available.}
\label{fig:zjk}
\end{figure}

\begin{figure}
\resizebox{0.48\textwidth}{!}{\includegraphics{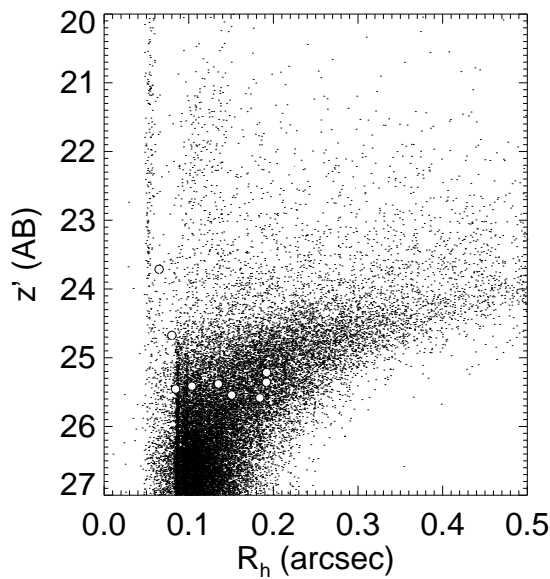}}
\caption{A magnitude-size diagram for our data illustrating the clearly
defined stellar locus at $R_h< 0.1''$. Candidate objects are marked
with circles --- 6 are clearly resolved, and 3 are compact (of which one
is likely to be a T-dwarf star, and another a putative AGN).}
\label{fig:magsize}
\end{figure}

Using our $(i'-z')$ colour-cut criterion alone may also select
elliptical galaxies at $z\sim 2$ (as shown by Figure~\ref{fig:tracks}),
the Extremely Red Object (ERO) population (e.g., Cimatti et al.\
2002). Another potential source of contamination are Galactic L- or
T-dwarf low-mass stars (e.g., Hawley et al.\ 2002). We can guard against
these lower-redshift interlopers by considering the near-infrared
colours. Deep $J$ and $K_s$ band observations of the ACS survey field
have been obtained with ISAAC on VLT, and reduced data from this ESO
Imaging Survey has also been publically released as part of the GOODS
programme\footnote{see {\tt
http://www.eso.org/science/goods/releases/20020408/}}. As
Figure~\ref{fig:zjk} illustrates near-infrared colours effectively
discriminate the ERO population from high redshift objects. Of the five
candidates satisfying our selection criteria for which such data is
available, one (object 2) is detected in F606W $v$-band ($v_{\rm
AB}=27.0\pm0,.2$) and has colours consistent with being an ERO
contaminant. All other candidates have $v_{\rm AB}>28.0$ (our
$3\,\sigma$ detection limit), although object 9 also has red near-IR
colours. Object 5 has a colours which suggests that it is likely to be a
low mass star: $(z'_{\rm AB}-J_{\rm Vega})_=3.12$ \& $(J-K_s)_{\rm
Vega}=0.2$ (an L- or T-dwarf).  Comparison with the sample in Hawley et
al.\ (2002) indicates a spectral type around T3. However, of the
remaining objects for which no reduced infrared data is yet available,
one is now confirmed as a high redshift object (see `Note Added in
Proof') and the others are all fully resolved as shown in
Figure~\ref{fig:magsize}. The contamination rate would thus appear to be
about 25 per cent.

\subsection{Limiting Star Formation Rates}
\label{sec:LimSFRs}

\begin{figure}
\resizebox{0.48\textwidth}{!}{\includegraphics{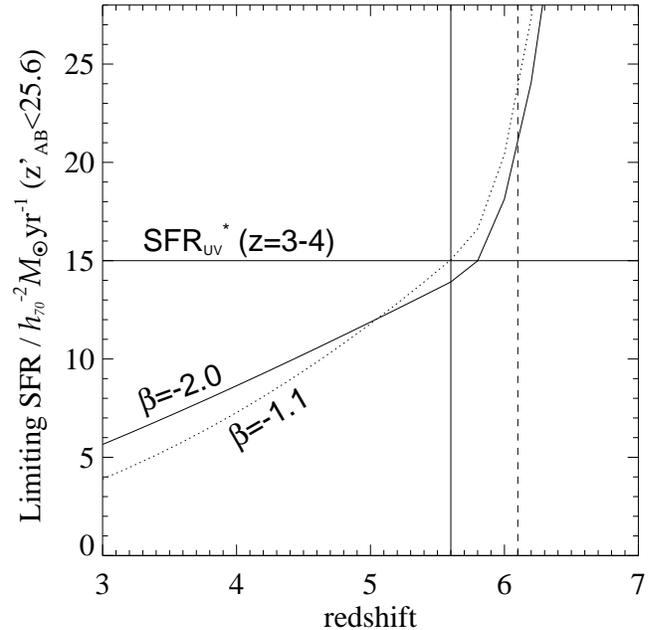}}
\caption{The limiting star formation rates reached as a function of
redshift for our catalogue with $z'_{\rm AB}<25.6$\,mag. The star formation
rates are inferred from the rest-frame UV continuum around 1500\,\AA\
(Madau, Pozzetti \& Dickinson 1998). We account for the
filter transmission, and the blanketting effects of the intervening
Lyman-$\alpha$ forest. We introduce small $k$-corrections from the
observed rest-wavelengths redshifted into the $z'$-band to 1500\,\AA :
the solid line assumes a spectral slope $\beta=-2.0$ (where
$f_{\lambda}\propto \lambda^{\beta}$) which is appropriate for an
unobscured starburst, and the dotted line has $\beta=-1.1$ (the slope for
mean reddening of the $z\approx 3$ Lyman break galaxies). The value of
${\rm SFR}^{*}_{\rm UV}=15\,h^{-2}_{70}\,M_{\odot}\,{\rm yr}^{-1}$ for
$z=3-4$ from Steidel et al.\ (1999) is the horizontal line. Our colour
selection should remove most $z<5.6$ galaxies (solid vertical line), and
our effective survey volume extends out to $z=6.1$ (dashed line).}
\label{fig:sfrlimits}
\end{figure}

We are sensitive to objects above $z \approx 5.6$ and in principle could
identify objects out to $z \approx 7.0$ where the Lyman-$\alpha$ line
leaves the $z'$ filter. The rapid falloff in sensitivity of the ACS
detectors above 9000\ \AA, however, combined with the transition of the
Lyman-$\alpha$ break through the $z'$ filter and the resulting
incomplete coverage of that filter, means our sensitivity to star
formation drops rapidly past $z\approx 6.0$, and renders us unlikely to
detect anything past $z\approx 6.5$ (Figure~\ref{fig:sfrlimits}). The
effect of luminosity-weighting of the redshift range also affects our
effective survey volume, and we quantify this in
Section~\ref{sec:survol}.

The observed luminosity function of Lyman break galaxies (LBGs) around
$\lambda_{\rm rest}=1500$\,\AA\ is $m^{*}_{R}=24.48$ at $\langle
z\rangle=3.04$ and $m^{*}_{I}=24.48$ at $\langle z\rangle =4.13$, from
Steidel et al.\ (1999), with a faint end slope $\alpha=-1.6$ and
normalization $\phi^{*}\approx 0.005\,h_{70}^3\,{\rm Mpc}^{-3}$. If
there is no evolution in the luminosity function from $z=3$ (as is found
to be the case at $z=4$) this would predict an apparent magnitude of
$z'_{\rm AB}=25.6$ for an $L^*_{\rm LBG}$ at $z\sim 6$. Thus our complete
catalogue of $z'_{\rm AB}<25.6$ would include galaxies down to $L^*_{\rm
LBG}$ at $z\sim 6$ if there is no evolution in the LBG luminosity
function.

The relation between the flux density in the rest-UV around $\approx
1500$\,\AA\ and the star formation rate (${\rm SFR}$ in $M_{\odot}\,{\rm
yr}^{-1}$) is given by $L_{\rm UV}=8\times 10^{27} {\rm SFR}\,{\rm
ergs\,s^{-1}\,Hz^{-1}}$ from Madau, Pozzetti \& Dickinson (1998) for a
Salpeter (1955) stellar initial mass function (IMF) with
$0.1\,M_{\odot}<M^{*}<125\,M_{\odot}$. This is comparable to the
relation derived from the models of Leitherer \& Heckman (1995).
However, if the Scalo (1986) IMF is used, the inferred star formation
rates are a factor of $\approx 2.5$ higher for a similar mass
range. From Steidel et al.\ (1999), the characteristic star formation
rate at the knee of the luminosity function, $L^*_{\rm LBG}$, is ${\rm
SFR}^{*}_{\rm UV}=15\,h^{-2}_{70}\,M_{\odot}\,{\rm yr}^{-1}$ for
$z=3-4$. Figure~\ref{fig:sfrlimits} shows our limiting star formation
rate as a function of redshift, inferred from the rest-frame UV flux in
the $z'$-filter. We account for the filter transmission, and the
blanketting effects of the intervening Lyman-$\alpha$ forest. We
introduce small $k$-corrections to $\lambda_{\rm rest}=1500$\,\AA\ from
the observed rest-wavelengths longward of Lyman-$\alpha$ redshifted into
the $z'$-band: we consider both a spectral slope $\beta=-2.0$ (where
$f_{\lambda}\propto \lambda^{\beta}$) which is appropriate for an
unobscured starburst (flat in $f_{\nu}$), and also a redder slope of
$\beta=-1.1$ (approriate for mean reddening of the $z\approx 3$ Lyman
break galaxies, Meurer et al.\ 1997). At limiting magnitude of
$z'_{\rm AB}=25.6$, we can detect unobscured star formation rates as low as
$15\,[16.5]\,h^{-2}_{70}\,M_{\odot}\,{\rm yr}^{-1}$ at $5.6<z<5.8$ and
$21\,[25]\,h^{-2}_{70}\,M_{\odot}\,{\rm yr}^{-1}$ at $z<6.1$ for
spectral slope $\beta=-2.0\,[-1.1]$. Our $(i'-z')_{\rm AB}>1.5$ colour cut
should remove most galaxies at $z<5.6$.

\subsection{Candidate Objects}
\label{sec:CandObjs}

\begin{figure}
\mbox{\resizebox{0.267\textwidth}{!}{\includegraphics{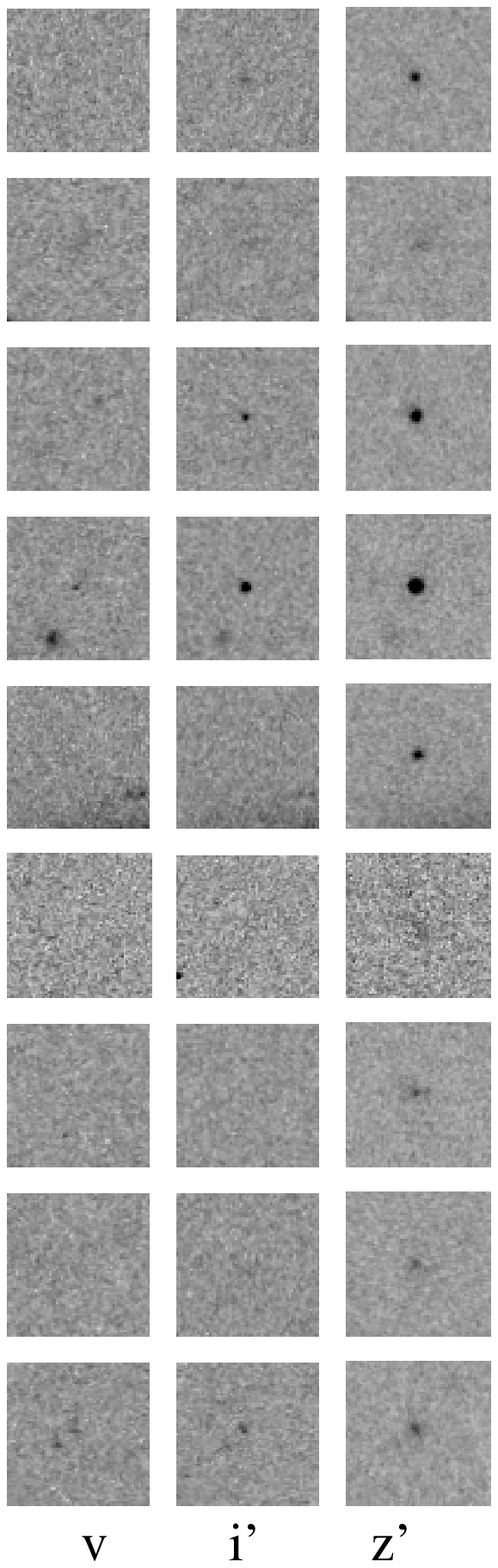}}
\resizebox{0.213\textwidth}{!}{\includegraphics{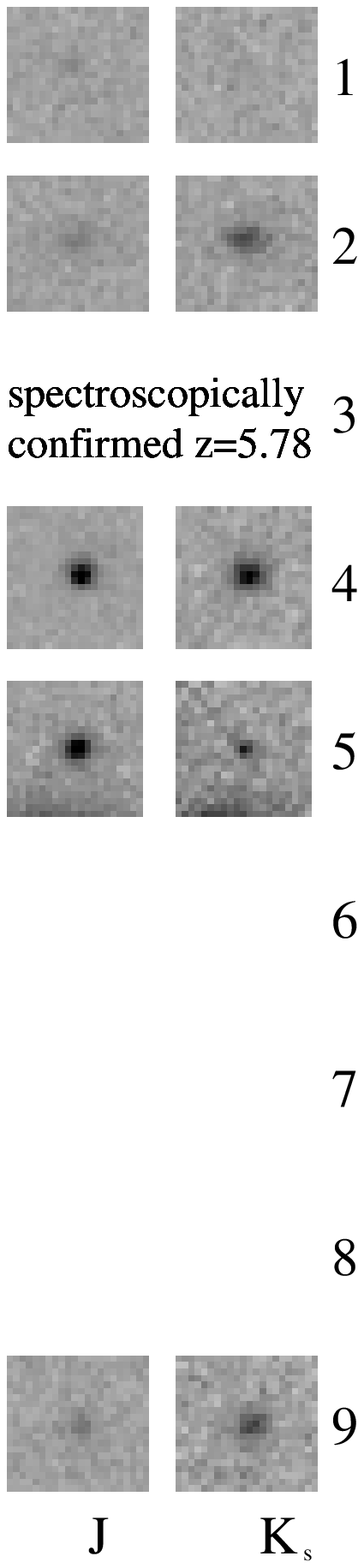}}}
\caption{Candidate $z>5.6$ galaxies selected to have $(i'-z')_{\rm AB}> 1.5$. 
The images shown are 3 arcseconds to a side and represent data from all three epochs of observation. Note that object 6 is not observed in epochs 1 and 3 due to the nature of the tiling scheme used.}
\label{fig:stamps}
\end{figure}

Using the selection criteria described above we have identified 7
candidate galaxies photometrically selected to lie at $z > 5.6$. These
(along with the probable star and ERO included for comparison) are
described in Table~1 and cut-out images of each object in the $z'$,
$i'$, $v$(F606W) and (where available) $J$ and $K_s$ bands are presented
in Figure~\ref{fig:stamps}.

As Figure~\ref{fig:magsize} illustrates most of our $(i'-z')_{\rm AB}>1.5$
objects are spatially resolved in the {\em HST}/ACS images (6 of the
9). Of the three compact objects, object 5 has the colours of an L- or
T-dwarf star. We note that object 4 is almost 1 magnitude brighter than
the next brightest object and lies close to the FWHM stellar locus, and
is thus likely to be an AGN.

\begin{table*}
\centering
\begin{tabular}{c|c|l|c|c|c|c|c|c|c|}
ID & RA & Declination & $z'_{\rm AB}$ & $i'_{\rm AB}$ & $(i'-z')_{\rm AB}$ &
$(z'_{\rm AB}-J_{\rm Vega})$ & $(J-K_s)_{\rm Vega}$ & $R_h$ &
SFR$^{z=5.5}_{\rm UV}$ \\
 & (J2000) & (J2000) & & & & & & arcsec &$h^{-2}_{70}\,M_{\odot}\,{\rm
yr}^{-1}$\\
\hline\hline
& & & & & & & & &\\
      1    &          3$^{\rm h}$32$^{\rm m}$40\fs00   &
$-$27\degr48\arcmin15\farcs0   &   25.41
$\pm$   0.07   &   26.91   $\pm$   0.22   &   1.50   $\pm$   0.23   &
0.90   $\pm$   0.15   &   0.72   $\pm$   0.25   &   0$\farcs$10    &   17.1   \\

      2    &          3$^{\rm h}$32$^{\rm m}$34\fs64   &
$-$27\degr47\arcmin20\farcs9   &   25.58
$\pm$   0.08   &   27.09   $\pm$   0.26   &   1.51   $\pm$   0.27   &
1.94   $\pm$   0.10   &   1.82   $\pm$   0.07   &   0$\farcs$18    &   14.7   \\

      3    &          3$^{\rm h}$32$^{\rm m}$25\fs59   &       
$-$27\degr55\arcmin48\farcs4   &   24.67
$\pm$   0.03   &   26.27   $\pm$   0.13   &   1.60   $\pm$   0.13   &
---    &     ---    &   0$\farcs$07    &   33.8   \\

      4    &          3$^{\rm h}$32$^{\rm m}$18\fs18   &       
$-$27\degr47\arcmin46\farcs5   &   23.71
$\pm$   0.01   &   25.26   $\pm$   0.05   &   1.54   $\pm$   0.05   &
1.72   $\pm$   0.02   &   1.33   $\pm$   0.03   &   0$\farcs$06    &   82.0   \\

      5    &          3$^{\rm h}$32$^{\rm m}$38\fs80   &       
$-$27\degr49\arcmin53\farcs6   & 25.45   
$\pm$   0.08   &    $>27.3$ $(3\,\sigma)$    &    $> 1.85$ $(3\,\sigma)$ &
   3.12   $\pm$   0.09   &   0.22   $\pm$   0.02   & 0$\farcs$07    &   16.5   \\

      6    &          3$^{\rm h}$32$^{\rm m}$06\fs48   &
$-$27\degr48\arcmin 46\farcs6   &
25.35   $\pm$   0.07   &   27.17   $\pm$   0.28   &   1.82   $\pm$   0.29
&     ---    &     ---    &   0$\farcs$19    &   18.1   \\

      7    &          3$^{\rm h}$32$^{\rm m}$33\fs19   &
$-$27\degr39\arcmin49\farcs2   &   25.38
$\pm$   0.07   &    $>27.3$ $(3\,\sigma)$   &    $> 1.92$  $(3\,\sigma)$ &
---    &     ---    &   0$\farcs$13    &   17.7   \\

      8    &          3$^{\rm h}$32$^{\rm m}$23\fs77   &
$-$27\degr40\arcmin37\farcs8   &   25.54
$\pm$   0.08   &    $>27.3$ $(3\,\sigma)$   &    $> 1.76$ $(3\,\sigma)$  &
---    &     ---    &   0$\farcs$15    &   15.2   \\

      9    &          3$^{\rm h}$32$^{\rm m}$18\fs18   &
$-$27\degr46\arcmin16\farcs1   &   25.21
$\pm$   0.06   &   26.75   $\pm$   0.19   &   1.54   $\pm$   0.20   &
1.88   $\pm$   0.11   &   2.15   $\pm$   0.10   &   0$\farcs$19    &   20.6   \\

& & & & & & & & & \\

\end{tabular}

\caption{Candidate $z > 5.6$ galaxies, satisfying our criteria
$z'_{AB}<25.6$\,mag and $(i'-z')_{\rm AB}>1.5$\,mag. All magnitudes are
aperture magnitudes within a 1\,arcsec diameter aperture, and the fluxes
have been averaged over the 3 available epochs of ACS data. For those
objects that were undetected in $i'$, the 3\,$\sigma$ limiting magnitude
of $i'_{\rm AB} = 27.3$ has been used to place a lower limit on the
$(i'-z')$ colour. `---' indicates that no data is available. The
half-light radius, $R_h$, is defined as the radius enclosing half the
flux. Stellar objects have $R_h\approx 0\farcs06$. Nominal star
formation rates (SFRs) are calculated assuming objects are placed in the
middle of our effective volume (a luminosity-weighted redshift of $z =
5.8$), and are based on the $k$-corrected 1500\,\AA\ rest-frame UV
continuum, assuming a spectral slope of $\beta=-2.0$; for a more
reddened slope of $\beta=-1.1$, the star formation rates are 10 per cent
higher. We use AB magnitudes for the {\em HST}/ACS images and Vega
magnitudes for the near-infrared $J$ and $K_s$ data; the conversions to
AB magnitudes are $(J_{\rm AB}-J_{\rm Vega})=0.981$ and $(K_{\rm
AB}-K_{\rm Vega})=1.843$ (Koornneef 1983).}
\label{tab}
\end{table*}

\begin{figure}
\resizebox{0.48\textwidth}{!}{\includegraphics{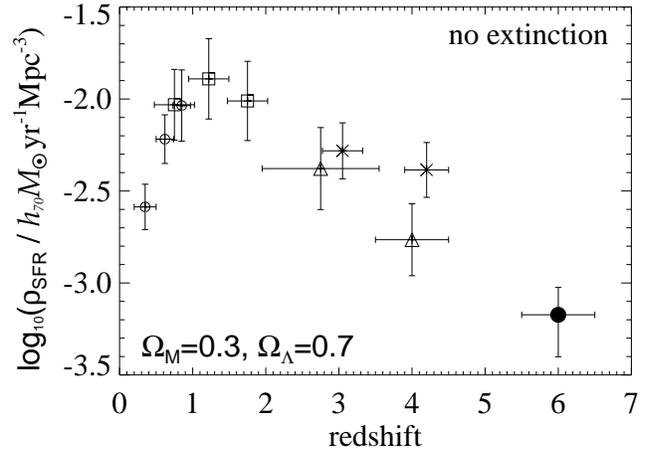}}
\caption{The `Madau-Lilly' plot illustrating the evolution of the
comoving volume-averaged star formation rate, reproduced from Steidel et
al.\ (1999) and recalculated for our cosmology and higher luminosity
limit of $1\,L^*$ (instead of $0.1\,L^*$).  The luminosity functions for
galaxies derived by Steidel et al.\ were used to recalculate these
values, assuming a slope of $\alpha=-1.6$ for $z>2$ and $\alpha=-1.3$
for $z<2$. Data from the CFRS survey of Lilly et al.\ (1996) are shown
as open circles; data from Connolly et al.\ (1997) are squares; the
HDF-North results from Madau et al.\ (1996) are triangles; and the Lyman
break galaxy work of Steidel et al.\ (1999) is plotted as crosses. The
result from this work is shown as a filled circle. Note that none of the
data-points on this figure are corrected for extinction.}
\label{fig:madau}
\end{figure}

\section{Global Star Formation at $z\sim 6$}
\label{sec:SFhistory}

\subsection{The Survey Volume}
\label{sec:survol}

Galaxies in the range $5.6<z<7.0$ would be selected by our colour-cut
provided they are sufficiently luminous. However, in practice we are
sensitive over a smaller range of redshifts, because galaxies become
progressively fainter at higher redshift, and so an ever smaller
fraction of them are able to satisfy the magnitude limit of
$z'_{\rm AB}=25.6$. There are three main reasons for this: 
\begin{enumerate}
\item The
selected galaxies lie primarily on the steep exponential cut-off of the
luminosity function (brighter than $L^{*}$ for a no-evolution model), so
small changes in the absolute magnitude limit with redshift will greatly
affect the number of galaxies brighter than the limit. 
\item
Another effect is the $k$-correction (Section~\ref{sec:LimSFRs}): as
redshift increases, the $z'$-band samples light in the rest-frame of the
galaxies at wavelengths that are increasingly far to the blue of
1500\,\AA , where the LBGs' luminosity function was calculated. This
$k$-correction makes higher redshift galaxies fainter (as LBGs are red,
with $\beta>-2.0$), and therefore less likely to satisfy the magnitude
limit. 
\item At redshifts $z>6$, Lyman-$\alpha$ absorption from
the forest enters the $z'$-band and makes galaxies fainter still, as
there is incomplete coverage of the filter by the continuum longward of
Lyman-$\alpha$.
\end{enumerate}

We have followed the approach of Steidel et al.\ in calculating the
effect of this luminosity bias on our sample of $z\sim 6$ LBGs. We
compute an effective survey volume using
\[
V_{\rm eff}(m)=\int dz\,p(m,z)\,\frac{dV}{dz}
\]
where $p(m,z)$ is the probability of detecting a galaxy at redshift $z$
and apparent $z'$ magnitude $m$, and $dz\,\frac{dV}{dz}$ is the comoving
volume per unit solid angle in a slice $dz$ at redshift $z$. We
integrate over the magnitude range we are sensitive to, and over the
redshift range $5.6<z<7.0$ from our colour selection. We calculate that
for a spectral slope of $\beta=-2.0$ (i.e., flat in $f_{\nu}$) the
effective comoving volume is 40 per cent the total volume in the range
$5.6<z<7.0$ (i.e., equivalent to $5.6<z<6.1$). For our 146\,arcmin$^{2}$
survey area (after excluding the unreliable regions close to the CCD
chip gaps) this is a comoving volume of $1.8\times
10^{5}\,h^{-3}_{70}\,{\rm Mpc}^{3}$. For a redder spectral slope of
$\beta=-1.1$ (the average of the $z=3$ LBGs, Meurer et al.\ 1997) the
effective volume is 36 per cent of the total.

Since our aim is to investigate the high-redshift progenitors of
present-day normal galaxies we must survey a comoving volume at $z\sim
6$ sufficient to encompass many $\sim L^*$ galaxies by the current
epoch, in order to reduce the effects of cosmic variance. From local
surveys, the number-density of $L^*$ galaxies is
$\phi^*=(0.0048\pm0.0006)\,h_{70}^3\,{\rm Mpc}^{-3}$ (Loveday et al.\
1992). Hence, the comoving volume occupied by the progenitors of a
present-day $L^*$ galaxy is $V^*\approx 200\,h_{70}^{3}\,{\rm Mpc}^3$.
Our $(i'-z')$ colour selection surveys an effective comoving volume of
$\approx 180,000\,h_{70}^{-3}\,{\rm Mpc}^3$. Thus in principle, assuming
the data reaches sufficient depth, we could detect the progenitors of
$\sim 1000$ present-day $L^*$ galaxies in the CDFS survey field.

Since significantly fewer objects than this have been identified in this
work we are able to conclude that at $z\sim6$ the ancestors of $L^*$
galaxies have continuous unobscured star formation rates lower than
$\approx 20\,h_{70}^{-2}\,M_{\odot}\,{\rm yr}^{-1}$ per galaxy over
$5.6<z<6.1$.

\subsection{The Star Formation Rate at $z\sim 6$: Lower limits and
Caveats} 
\label{sec:caveats}

Our selection criteria have identified 6 candidate star-forming galaxies
with $z>$5.6, (excluding the probable star, the ERO and the potential
AGN with nominal SFRs of 17, 15 and 82 $h_{70}^{-2}\,M_{\odot}\,{\rm
yr}^{-1}$ respectively). These have an integrated star formation rate of
$(122\pm 50)\,h_{70}^{-2}\,M_{\odot}\,{\rm yr}^{-1}$ (where the error is
based on Poisson statistics), giving a comoving volume-averaged star
formation density between $z = 5.6$ and $z=6.1$ of $\rho_{\rm
SFR}=(6.7\pm 2.7) \times 10^{-4}\,h_{70}\,M_{\odot}\,{\rm yr}^{-1}\,{\rm
Mpc}^{-3}$.

This value is compared to global star-formation rates found at lower
redshift in Figure~\ref{fig:madau}, reproduced from Steidel et al.\
(1999) and recalculated for our $\Lambda$ cosmology and higher limiting
luminosity. It appears that the comoving star formation rate (based on
the bright end of the UV luminosity function) was 6 times {\em less} at
$z\sim 6$ than at $z=3-4$. The interpretation of this result,
however, depends on several considerations.

Firstly, the star formation densities plotted have not been corrected
for dust extinction within the galaxies in question. Although this will
reduce the observed UV flux, and hence the implied star formation rate,
the method used to select galaxies (the Lyman break technique) has been
consistently applied at all redshifts above $z\approx 2$. Thus
equivalent populations should be observed at all redshifts and the
resulting dust correction will alter the normalization of all the data
points in the same way if there is no evolution in the extinction.

Secondly, previous authors have integrated the star formation density
over the luminosity function for their objects down to star formation
rates of 0.1\,$L^{*}$ even if these SFRs have not been observed. We
prefer to plot what we actually observe and not to extrapolate below the
limit of our observations. If the luminosity function is extended down
to $0.1\,L^*$ then the global SFR is increased by a factor of $\sim5$ for
$\alpha=-1.6$.

Furthermore, Lanzetta et al.\ (2002) note that at high redshifts the
effects of cosmological surface dimming of objects can lead to failure
to detect dispersed regions of low star formation and thus to an
underestimation of the total star formation density when calculations
are based purely on compact and intense regions of high star formation.

Since these potential corrections to our data are expected to increase
the volume-averaged star formation density, the value quoted above
should be considered a firm lower limit rather than an absolute
determination. Combining this work with planned deeper {\em HST}/ACS
observations in the $i'$- and $z'$-filters, such as the Ultra Deep
Field, will enable fainter magnitudes to be probed and allow the slope
of the luminosity function to be determined (e.g., Yan et al.\ 2003),
further constraining the integrated star formation rate and associated
contribution to the UV background.

\section{Conclusions}
\label{sec:conclusion}

We have determined the space density of UV-luminous star-burst galaxies
at $z\sim 6$ using deep {\em HST} ACS SDSS-$i'$ (F775W) and SDSS-$z'$
(F850LP) and VLT ISAAC $J$- and $K_s$- band imaging of the Chandra Deep
Field South from the GOODS survey. The Lyman Break technique was used to
identify 7 candidate high redshift objects, of which one may be an
AGN. We reach an unobscured star formation rate of $\approx
15\,h_{70}^{-2}\,M_{\odot}\,{\rm yr}^{-1}$ at $z\sim 6$, equivalent to
$L^*$ for the Lyman break population at $z\sim 3-4$ over a comoving
volume of $\approx 1.8\times10^5\,h^{-3}_{70}\,{\rm Mpc}^{3}$ -- which
should encompass the progenitors of about a thousand $L^*$ galaxies at
the current epoch. It appears that the comoving star formation rate
(based on the bright end of the UV luminosity function above $L^{*}$)
was 6 times {\em less} at $z\sim 6$ than at $z=3-4$. These results have
used to determine a lower bound on the integrated volume averaged global
star formation rate at $z\sim 6$. Hence we have been able to extrapolate
the star formation history of the Universe to earlier epochs than the
existing Madau-Lilly diagram. These observations provide an important
lower limit on contribution of galaxies to the UV background over a
redshift range that some recent observational evidence suggests is the
epoch of reionization.

\subsection*{Note Added in Proof}
\label{sec:note}

We have recently obtained spectroscopy on a subset of the objects
described in this paper using the DEIMOS instrument on the Keck II
telescope, in collaboration with Prof.\ Richard Ellis (Caltech) and Dr.\
Patrick McCarthy (OCIW).  While analysis of this data is ongoing, one of
the objects (Object 3) shows Ly-$\alpha$ in emission at a redshift of
$5.78$, and we report this spectrum in Bunker et al.\ (2003, {\tt
astro-ph/0302401}) submitted to MNRAS (Letters).

\subsection*{ACKNOWLEDGMENTS}

ERS acknowledges a Particle Physics and Astronomy Research Council
(PPARC) studentship supporting this study. We are extremely grateful to
the referee, Kurt Adelberger, for his constructive comments
(particularly on the issue of the luminosity-weighted effective volume
of the survey). We thank Richard Ellis, Andrew Firth, Simon Hodgkin,
Ofer Lahav, Pat McCarthy and Rob Sharp for useful discussions. This
paper is based on observations made with the NASA/ESA Hubble Space
Telescope, obtained from the Data Archive at the Space Telescope Science
Institute, which is operated by the Association of Universities for
Research in Astronomy, Inc., under NASA contract NAS 5-26555. These
observations are associated with proposals \#9425\,\&\,9583 (the GOODS
public imaging survey). We are grateful to the GOODS team for making
their reduced images public -- a very useful resource. The archival
GOODS/EIS infrared images are based on observations collected at the
European Southern Observatory, Chile, as part of the ESO Large Programme
LP168.A-0485(A) (PI: C.\ Cesarsky), and ESO programmes 64.O-0643,
66.A-0572 and 68.A-0544 (PI: E.\ Giallongo).

\bsp


\begin{thebibliography}{}
\bibitem[\protect\citename{Becker et al.\ } 2001]{be01}
Becker R.~H. et al., 2001, AJ, 122, 2850

\bibitem[\protect\citename{Bertin \& Arnouts} 1996]{ba96} 
Bertin E., Arnouts S., 1996, A\&AS, 117, 393

\bibitem[\protect\citename{Bunker et al.\ } 2003]{bu03} 
Bunker A.~J., Stanway E.~R., Ellis R.~S., McMahon R.~G., McCarthy P.~J.,
2003, preprint (astro-ph/0302401), submitted to MNRAS (Letters)

\bibitem[\protect\citename{Cimatti et al.\ } 2002]{ci02}
Cimatti A. et al., 2002, A\&A, 381, L68

\bibitem[\protect\citename{Coleman, Wu \& Weedman } 1980]{co80}
Coleman G.~D., Wu C.-C., Weedman D.~W., 1980, ApJS, 43, 393

\bibitem[\protect\citename{Connolly et al.\ } 1997]{co97}
Connolly A.~J., Szalay A.~S., Dickinson M., SubbaRao M.~U., Brunner
R.~J., 1997, ApJ, 486, L11

\bibitem[\protect\citename{Cowie et al.\ }1996]{co96}
Cowie L.~L., Songaila A., Hu E.~M., Cohen J.~G., 1996, 
AJ, 112, 839 

\bibitem[\protect\citename{Dawson et al.\ } 2001]{da01}
Dawson S., Stern D., Bunker A., Spinrad H., Dey A., 2001, AJ, 122, 598

\bibitem[\protect\citename{Dickinson et al.\ } 2000] {da00}
Dickinson M. et al., 2000, ApJ, 531, 624

\bibitem[\protect\citename{Dickinson \& Giavalisco } 2002] {dg02} 
Dickinson M., Giavalisco M., 2002, in Bender R., Renzini A., eds., ESO
Astrophysics Symposia Series, The Mass of Galaxies at Low and High
Redshift, Springer-Verlag, Berlin, p.~324

\bibitem[\protect\citename{Fan et al.\ } 2000] {fa00}
Fan X. et al., 2000, AJ, 122, 2833

\bibitem[\protect\citename{Ford et al.\ } 2002]{fo02} 
Ford H.~C. et al., 2002, BAAS, 200.2401

\bibitem[\protect\citename{Fruchter \& Hook } 2002]{fr02}
Fruchter A., Hook R., 2002, PASP, 114, 144

\bibitem[\protect\citename{Hawley et al.\ } 2002]{ha02}
Hawley S.~L. et al., 2002, AJ, 123, 3409

\bibitem[\protect\citename{Hu \& McMahon } 1996]{hm96}
Hu E.~M., McMahon R.~G., 1996, 
Nature, 382, 281

\bibitem[\protect\citename{Hu, McMahon \& Cowie } 1999]{hu99}
Hu E.~M., McMahon R.~G., Cowie L.~L., 1999, 
ApJ, 522, L9

\bibitem[\protect\citename{Hu et al.\ } 2002]{hu02}
Hu E.~M., Cowie L.~L., McMahon R.~G., Capak P., Iwamuro F., Kneib J.-P., Maihara T., Motohara K., 2002,
ApJ, 568, L75

\bibitem[\protect\citename{Kodaira et al.\ } 2003]{ko03}
Kodaira K. et al., 2003, preprint, (astro-ph/0301096)

\bibitem[\protect\citename{Kogut et al.\ } 2003]{kog03}
Kogut A., et al., 2003, preprint (astro-ph/0302213)


\bibitem[\protect\citename{Koornneef }1983]{ko83}
Koornneef J., 1983, A\& A, 128, 84

\bibitem[\protect\citename{Lanzetta et al.\ } 2002]{la02} 
 Lanzetta K.~M., Yahata N., Pascarelle S., Chen H.-W., Fern\'andez-Soto A., 
2002, ApJ, 570, 492

\bibitem[\protect\citename{Lehnert \& Bremer } 2003]{le03}
Lehnert M.~D., Bremer M., 2003, preprint (astro-ph/0212431)

\bibitem[\protect\citename{Leitherer \& Heckman } 1995]{le95}
Leitherer C., Heckman T.~M., 1995, ApJS, 96, 9

\bibitem[\protect\citename{Lilly et al.\ } 1996]{li96}
Lilly S.~J., Le F\`{e}vre O., Hammer F., Crampton D., 1996,
ApJ, 460, L1 

\bibitem[\protect\citename{Loveday et al.\ } 1992]{lo92}
Loveday J., Peterson B. A., Efstathiou G., Maddox S. J., 1992, 
ApJ, 390, 338 

\bibitem[\protect\citename{Madau et al.\ } 1996]{ma96}
Madau P., Ferguson
H.~C., Dickinson M.~E., Giavalisco M., Steidel C.~C., Fruchter A., 1996,
MNRAS, 283, 1388

\bibitem[\protect\citename{Madau, Pozzetti \& Dickinson } 1998]{ma98}
Madau P., Pozzetti L., Dickinson M., 1998, ApJ, 498, 106

\bibitem[\protect\citename{Meurer et al.\ } 1997]{me97}
Meurer G,~R., Heckman T.~M., Lehnert M.~D., Leitherer C., Lowenthal J.,
1997, AJ, 114, 54

\bibitem[\protect\citename{Oke \& Gunn} 1983] {og83}
Oke J.~B., Gunn J.~E., 1983, ApJ, 266, 713

\bibitem[\protect\citename{Salpeter } 1955]{sa55}
Salpeter E.~E., 1955, ApJ, 121, 161

\bibitem[\protect\citename{Scalo } 1986]{sc86}
Scalo J.~M., 1986, Fund.\ Cosmic Phys., 11, 1

\bibitem[\protect\citename{Schlegel, Finkbeiner \& Davis} 1998] {sfg98}
Schlegel D.~J., Finkbeiner D.~P., Davis M., 1998,
ApJ, 500, 525 

\bibitem[\protect\citename{Spinrad et al.\ } 1998]{sp98}
Spinrad H., Stern D., Bunker A., Dey A., Lanzetta K., Yahil A., Pascarelle S., Fern\'andez-Soto A., 1998,
AJ, 116, 2617 

\bibitem[\protect\citename{Steidel, Pettini \& Hamilton }1995]{st95}
Steidel C.~C. Pettini M., Hamilton D., 1995, AJ, 110, 2519

\bibitem[\protect\citename{Steidel et al.\ } 1996]{st96}
Steidel C.~C., Giavalisco M., Pettini M., Dickinson M.~E., 
Adelberger K.~L., 1996, ApJ, 462, L17

\bibitem[\protect\citename{Steidel et al.\ } 1999]{st99}
Steidel C.~C., Adelberger K.~L., Giavalisco M., Dickinson M.~E., 
Pettini M., 1999, ApJ, 519, 1

\bibitem[\protect\citename{Thompson, Weymann \& Storrie-Lombardi } 2001]{tws01}
Thompson R.~I., Weymann R.~J., Storrie-Lombardi L.~J., 2001, ApJ, 546, 694 

\bibitem[\protect\citename{Weymann et al.\ } 1998]{we98}
Weymann R.~J., Stern D., Bunker A.~J., Spinrad H., Chaffee F.~H., 
Thompson R.~I., Storrie-Lombardi L.~J., 1998, ApJ, 505, L95

\bibitem[\protect\citename{Yan, Windhorst \& Cohen } 2003]{yan03}
Yan H., Windhorst R.~A., Cohen S., 2003, preprint (astro-ph/0212179)

\end{thebibliography}
\end{document}